\documentclass[twocolumn]{aastex6}
\usepackage{amsmath}    
\usepackage{mathtools}
\usepackage[english]{babel}
\usepackage{graphicx}

\begin{document}


\title{Collisionless electron-ion shocks in relativistic unmagnetized jet-ambient interactions: Non-thermal electron injection by double layer}


\author{Kazem Ardaneh\altaffilmark{1} and Dongsheng Cai}
\affil{Department of Computer Science, University of Tsukuba, Tenoudai 1-1-1, Tsukuba science city, Ibaraki 305-8573, Japan}
\and
\author{Ken-Ichi Nishikawa}
\affil{Department of Physics, University of Alabama in Huntsville, ZP12, Huntsville, AL 35805, USA}
\altaffiltext{1}{kazem.arrdaneh@gmail.com}
\begin{abstract}

The course of non-thermal electron ejection in relativistic unmagnetized electron-ion shocks is investigated by performing self-consistent particle-in-cell simulations. The shocks are excited through the injection of relativistic jet into ambient plasma, leading to two distinct shocks (named as the trailing shock and leading shock) and a contact discontinuity. The Weibel-like instabilities heat the electrons up to approximately half of ion kinetic energy. The double layers formed in the trailing and leading edges  then accelerated the electrons by the ion kinetic energy. The electron distribution function in the leading edge shows a clear non-thermal power-law tail which contains $\sim1\%$ of electrons and $\sim8\%$ of electron energy. Its power-law index is -2.6. The acceleration efficiency is $\sim23\%$ by number and $\sim50\%$ by energy and the power-law index is -1.8 for electron distribution function in the trailing edge. The effect of the dimensionality is examined by comparing results of 3D simulation with 2D ones. It exhibits that the electron acceleration is more efficient in 2D.

\end{abstract}

\keywords{acceleration of particles --- galaxies: jets --- cosmic rays --- plasmas --- shock waves}



\section{Introduction}\label{Introduction}

Tightly collimated streams of plasma with speeds close to the speed of light, commonly referred to as relativistic jets, are present in a variety of astrophysical objects, e.g., pulsar wind nebulae (PWNe) powered by the relativistic wind of pulsars, gamma-ray bursts (GRBs) from the death of massive stars or compact star mergers, and active galactic nuclei (AGNs) at the center of galaxies. The relativistic jets propagate through the ambient medium and excite the double shock structures afterwards. Acceleration of particles is ubiquitous in the astrophysical shocks \citep{koy95,eri11,mas13}. Non-thermal emission from these environments is usually considered as synchrotron or inverse Compton radiation from a power-law distribution of electrons accelerated at shock sides \citep{tau12}. Due to the lack of a perfectly self-consistent theory of particle acceleration in relativistic shocks, the power-law index and the acceleration efficiency, i.e., the fraction of particles and energy in the non-thermal tail, are usually compared with the observations.

Charged particles may be accelerated via first-order Fermi acceleration (or diffusive shock acceleration, DSA) in the collisionless shocks. In DSA, particles diffuse back and forth across the shock front and gain energy by scattering from the magnetohydrodynamics waves \citep{blan78,bel78,dru83,blan87,bel13}. However, DSA needs a seed population of particles with energies well in excess of the thermal ones, because only these particles are capable for multiple crossing the shock front and effective scattering by magnetic turbulences. However, it is not apparent how the electrons can reach the threshold energy of DSA. It demands their kinetic energies be comparable to those of the ions. This is known as the electron injection problem \citep{bal13}.

In the case of magnetized upstream, the injection of electrons is thought to be directly associated with the background motional electric field {\boldmath${E}_0=-{\beta}_0\times{B}_0$}. They may gain energy from the motional electric field while they gyrate--surf around the shock front. Based on the barrier that reflects the electrons toward the upstream, thus capable them for repeatedly energizations, this process is known with distinct names. If the reflecting barrier has a magnetic source, e.g., gradient of the magnetic field at the leading edge of the shock, the acceleration mechanism is named shock drift acceleration or SDA \citep{che75,web83,beg90,par12,par13,guo14}. If the barrier has an electrostatic source, e.g., the electrostatic solitary waves appeared at the leading edge of the shock by Buneman instability \citep{bun58}, the process is called shock surfing acceleration or SSA \citep{lee96,hos02,sha03,ama09,mat12}. Basically, the SSA process acts only in the electron-ion shocks, because electrostatic barrier would not be generated if the species have the same inertia. Magnetization parameter, obliquity angle of the upstream magnetic field with respect to the shock direction of propagation, and bulk Lorentz factor of the incoming stream may also play significant role in determining the responsible process for particle acceleration. 

An interesting question is: ``how does the electron ejection operate in unmagnetized electron-ion shocks?''. Due to the lack of upstream motional electric field, we expect a process other than SDA and SSA. Particle-in-cell (PIC) simulations provide a self-consistent description of particle acceleration in collisionless shocks. Our works have been allocated to the large scale PIC simulations of electron injection in unmagnetized relativistic electron-ion shocks. In PIC simulations, the shock waves are usually excited by the so-called injection approach \citep{hos01,hos02,spi8a,spi8b,ama09,mar09,sir11,sir13,guo14}. Using this approach, a high-speed plasma stream is launched from one end of the computational grid and reflected from a rigid boundary at the opposite end. Subsequently, a shock is excited due to the interactions between the incoming and the reflected streams. Although this method reduces by one-half the number of calculations, it has some disadvantages as well. In this method, the reverse and forward shock are degenerated (not distinguishable) and the simulations are limited to two identical counter-streaming beams. However, we are interested in asymmetric jet-ambient interaction, i.e., the interaction of plasmas with different properties that results in two different shocks, trailing shock (TS) and leading shock (LS), and a contact discontinuity (CD). 

In the present work, we have performed a 3D PIC simulation where collisionless double shock is created by an unmagnetized relativistic jet propagating into an unmagnetized ambient plasma. In contrast to the injection method, our asymmetric jet-ambient model is more realistic since it avoids an infinitely sharp CD and permits us to appropriately explore the dynamics of the TS and LS for different jet-ambient parameters. Beam-plasma (or jet-ambient) systems are susceptible to several instabilities, e.g., electrostatic two-stream or Buneman modes \citep{bun58}, and electromagnetic filamentation \citep{fri59} or Weibel \citep{wei59} modes. Therefore, the unstable spectrum is not less than 2D. Which of these modes will dominate highly relies on the system parameters \citep{bre09}. This undoubtedly clears a demand for studies using a method like ours \citep{nish03,nish05,nish09,nish16,ard14,ard15,cho14}, or using asymmetric counter-streaming beams \citep{niem12,wie16}, cause the most unstable modes excited in various setups can generate the totally different shock waves.

Our paper is dedicated to answer five questions: First, how does the double shock structure form in the unmagnetized jet-ambient interactions? Second, the shocks are characterized by magnetic or electrostatic forces? Third, what are the main processes responsible for electron injection? Forth, what is the resulting electron distribution function? Fifth, what is the effect of the dimensionality?

The simulation model and parameters setup are described in Section \ref{Simulation model and parameters setup}. The results of the simulations are presented in Section \ref{Simulation results}. We conclude with a summary in Section \ref{Summary and conclusions}.

\section{Simulation model and parameters setup}\label{Simulation model and parameters setup}
\begin{figure*}[tbph]
\begin{center}
\includegraphics[scale=0.4]{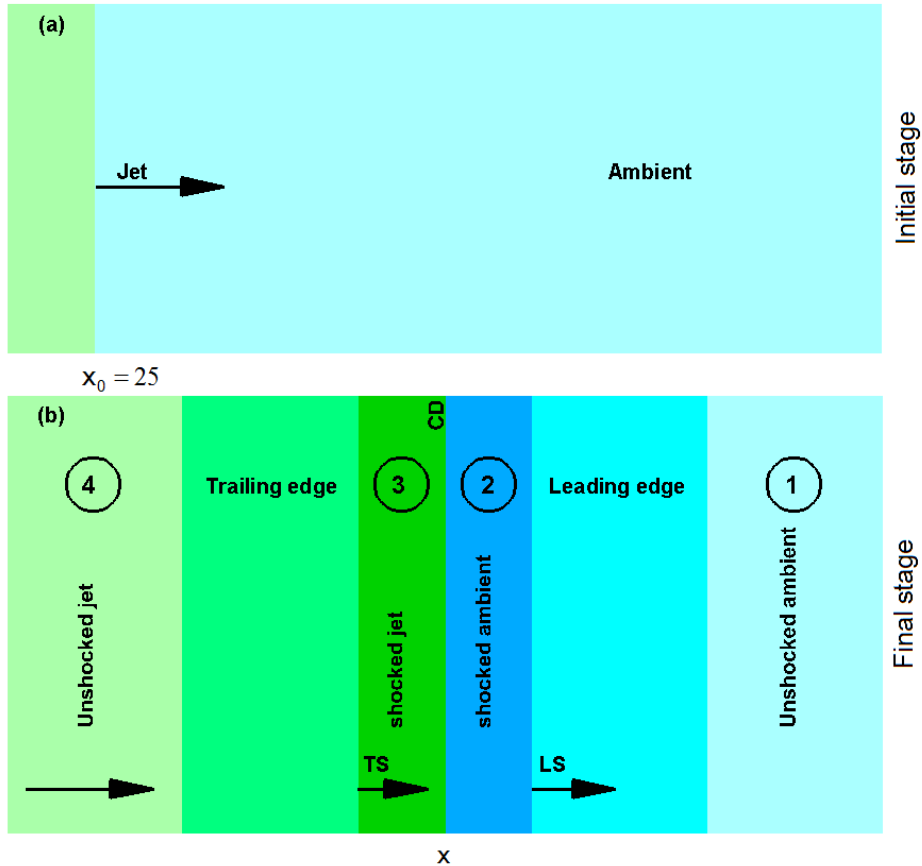}
\caption{An illustration of the jet-ambient interaction showing (a) a particles jet injecting into an ambient plasma and (b) the resulting double shock structure. Naming the shocks is according to \citet{nish09}. }
\label{collision}
\end{center}
\end{figure*}

In our works, an unmagnetized particles jet is injected into an unmagnetized ambient plasma \citep{nish03,nish05,nish09,nish16,ard14,ard15,cho14}. Finally, a double shock structure is formed resemble what is schematically illustrated in Figure \ref{collision}. Deceleration of the jet stream by the magnetic fluctuations (excited in the beam-plasma interactions) results in a CD and two shock waves that divide the jet and ambient plasmas into four regions: (1) unshocked ambient, (2) shocked ambient, (3) shocked jet, and (4) unshocked jet. Henceforward, subscripts 1, 2, 3, and 4 direct to the unshocked ambient, shocked ambient, shocked jet, and unshocked jet, respectively. Quantities with a single index $\varrho_{\rm i}$ indicate the value of quantities $\varrho$ in region $i$ in rest frame $i$ and quantities with double indices $\varrho_{\rm ij}$ show the value of quantities $\varrho$ in region $i$ as seen in rest frame $j$.

The code employed in the present work is an adopted version of the relativistic electromagnetic particle code TRISTAN \citep{bun93} with MPI-based parallelization \citep{niem08}. A series of test simulations are already performed to establish a numerical model which best conserves energy and minimizes numerical self-heating. In the PIC simulations including a cold relativistic plasma beam, a numerical heating instability arises when the beam propagates large distances over the numerical grid. The instability is a combination of grid-Cherenkov instability and spurious plasma oscillations \citep{ die06}. The latter oscillations are usually excited by coupling between a sideband of the beam mode with the electromagnetic mode. The beam mode has a physical phase speed $\omega/k=v_{b}$, where $v_{b}$ is the beam velocity. The beam interaction with the numerical grid, probably through a finite grid instability \citep{ bir91}, excites artificial sidebands that are separated from the beam mode by the frequency modulus $\Delta\omega =2n\pi v_{b}/\Delta$, where $2\pi v_{b}/\Delta$ is the grid crossing frequency. One of these sidebands may couple to the electromagnetic mode and results in the artificial obliquely propagating waves that are observed in the PIC simulations \citep{ die06}. The growth rate of these waves can be reduced by using a higher-order numerical scheme \citep{yee66,die06}. Here, the numerical instability is diminished by means of the fourth-order solver for Maxwell's curl equations and a weak Friedman filter as presented in \citet{gre04}.

The simulation is performed using a computational gird with $(\ell_{\rm x}, \ell_{\rm y}, \ell_{\rm z})=(8005, 245, 245)$, grid size: $\Delta x=\Delta y=\Delta z=1$. There are six particles per cell per species for the ambient plasma ($\simeq$ three billions particles per species). The density ratio between the jet and the ambient is $5/3$. The simulation frame of reference is the ambient, in which the jet plasma moves to the right in positive $x$-direction with bulk speed $\beta_{41}=0.995$ (bulk Lorentz factor $\Gamma_{41}=10$). The jet fills the whole computational box in the $yz$-plane and is injected continuously at $x_0 = 25$. The jet plasma is injected with energy distribution in the jet rest frame given by a 3D Maxwell-J\"{u}ttner distribution $f(\gamma_4) \propto \gamma_4^2\beta_4\exp(-\gamma_4/\theta_4)$ and thermal spread $\theta_4=(K_{\rm B}T_{\rm e}/m_{\rm e}c^2)_4=0.092$ (relativistically hot, $\beta_{\rm th4}=0.4$). In the ambient medium, the electrons have a thermal spread $\theta_1=(K_{\rm B}T_{\rm e}/m_{\rm e}c^2)_1=12.5\times10^{-4}$. In both plasmas, the ions are in thermal equilibrium with the electrons. The $m_{\rm i}/m_{\rm e}$ mass ratio used is 16. The system is numerically resolved with five grid cells per electron skin depth, $\lambda_{\rm ce}=5$, and $\Delta t=0.01\omega_{\rm pe}^{-1}$, where $\Delta t$ and $\omega_{\rm pe}$ are the time step and the electron plasma frequency, respectively. The surfaces at $x_{\min}$ and $x_{\max}$ are rigid reflecting boundaries for the ambient particles, while they are open boundaries for the jet particles. These surfaces are radiating boundaries for the fields based on Lindman's method \citep{lin75}. Periodic boundary conditions are applied at all other boundaries for both particles and fields. Hereafter, time is normalized to $\omega_{\rm pe}^{-1}$, space to the $\lambda_{\rm ce}$, particle momentum for species $s$ to the corresponding $m_{\rm s}c$ (e: electron and i: ion), and density to the unshocked ambient density, $n_1$. Furthermore, the position $x$ is measured from $x_0$.

For the described setup, according to the hydrodynamic jump conditions for jet-ambient interactions \citep{zha05,nish09,ard15}, the theoretical predictions for the LS and TS parameters under the adiabatic index $\tilde{\Gamma}=4/3$ are summarized in Table \ref{dsh}.

\begin{table*}[tbph]
\begin{center}
\caption{Parameters of the formed double shock structure.\label{dsh}}
\begin{tabular}{ |p{4cm}||p{4cm}||p{4cm}|  }
\hline
 \multicolumn{3}{|c|}{Parameters of the LS}\\
 \hline
Parameter & In region (1) & In region (2)\\
 \hline
 $\gamma_{\rm ls}$   & $\gamma_{\rm ls1}=1.91$    &$\gamma_{\rm ls2}=1.01$\\
 $\beta_{\rm ls}$&   $\beta_{\rm ls1}=0.85$  & $\beta_{\rm ls2}$=0.17 \\
 $n_2/n_1$ &$n_{21}/n_1=16.0$ & $n_{2}/n_{12}=5.8$\\
 \hline
 \multicolumn{3}{|c|}{Parameters of the TS}\\
 \hline
Parameter & In region (1) & In region (3)\\
 \hline
 $\gamma_{\rm ts}$   & $\gamma_{\rm ts1}=1.38$    &$\gamma_{\rm ts3}=1.03$\\
 $\beta_{\rm ts}$&   $\beta_{\rm ts1}=0.68$  & $\beta_{\rm ts3}$=-0.25 \\
 $n_3/n_4$ &$n_{31}/n_{41}=2.8$ & $n_{3}/n_{43}=4.9$\\
 \hline
\end{tabular}
\end{center}
\end{table*}

\section{Simulation results}\label{Simulation results}

The jet-ambient interactions include growth of the oblique instability \citep{bre10}, and the generation of magnetic fields which decelerate the jet stream and consequently form a double shock structure. At late times the particles are effectively heated, and accelerated. This section aims to explain the scenario in more detail.

\subsection{Formation of the CD}\label{Formation of the CD}

When the particles jet interacts with the ambient plasma, the distribution of particles is extremely anisotropic and is susceptible to several instabilities, e.g., electrostatic modes (two-stream or Buneman instabilities), and electromagnetic modes (filamentation or Weibel instabilities). Depending on the jet-to-ambient density ratio, jet and ambient temperatures, and jet drift velocity, two-stream, filamentation, or oblique modes will dominate the linear phase. Whereas perturbations parallel and normal to the jet stream are potentially present, the instability propagates obliquely.

\begin{figure}[h]
\begin{center}
\includegraphics[scale=0.3]{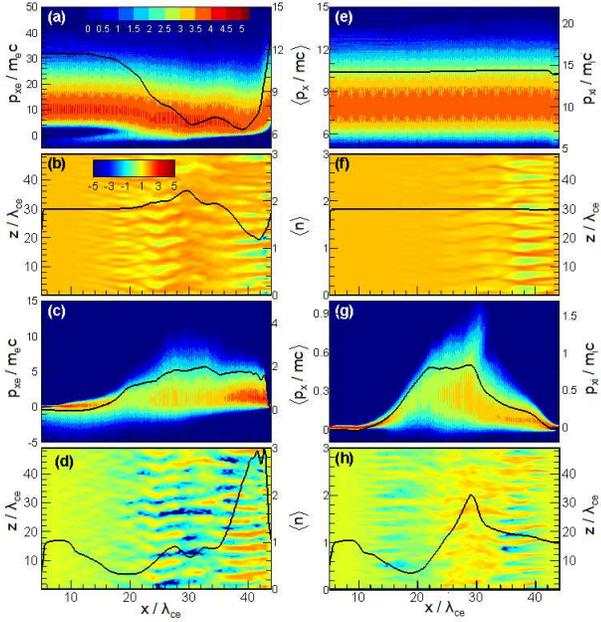}
\caption{Structure of the jet-ambient interaction at time $t=40\omega_{\rm pe}^{-1}$ when the fastest jet ions reach $x=45\lambda_{\rm ce}$. The longitudinal phase-space distribution and density in log scale are displayed for: jet electrons in panels (a) and (b), ambient electrons in panels (c) and (d), jet ions in panels (e) and (f), and ambient ions in panels (g) and (h). Over-plotted line in panel (a), (c), (e), and (g), shows the average momentum in $x$-direction. Over-plotted line in panel (b), (d), (f), and (h), shows the transversely averaged (in $yz$-plane) density normalized to the density in the unshocked ambient. In panels (a), (c), (e), and (g), the phase-space distributions are expressed in $\log[N(x,p_{\rm x})]$.}
\label{CDF}
\end{center}
\end{figure}

The jet electrons are rapidly decelerated when interact with ambient particles to form electron current filaments in both jet and ambient plasmas (Figure \ref{CDF}a). As a result, the density of the jet electron increases from $n_{41}/n_1=1.7$ to $n_{41}/n_1\simeq2.2$ just behind the jet front (Figure \ref{CDF}b). On other hand, ambient electrons become swept by the incoming jet stream (Figure \ref{CDF}c) and the density of the ambient electron is increased by a factor of three near the jet front (Figure \ref{CDF}d). In this stage (about $t=40\omega_{\rm pe}^{-1}$), a CD is formed around $x=36\lambda_{\rm ce}$ that separate decelerated jet electrons pile from the accelerated ambient electrons pile. The decelerated jet electrons become mainly confined to the left side of CD and pile up in this region. However, due to the CD formation, the accelerated ambient electrons are dominantly confined to the right side of the CD and they pile up due to the continuous sweeping by the jet inflow. Once trapped in the left side of CD, the jet electron populations commence heating. 

Due to ion larger inertia, the jet ions are able to penetrate deeper into the ambient plasma without significant deceleration (Figures \ref{CDF}e and \ref{CDF}f) and ambient ions are present in deeper length of the jet stream (Figures \ref{CDF}g and \ref{CDF}h). Therefore, a certain fraction of both ion populations (jet and ambient) is present in each other before the CD be fully formed. These fractions form a separate population on the two sides of the CD. Each of this population is affected by another plasma medium (jet or ambient) and is reflected back towards the CD. For the ambient fraction, since the ions have no means to either pass through the CD or escape from the continuous flow of jet particles, they are trapped in the left side of CD and will eventually become part of the TS population. This population is visible in the ambient ion phase-space and density plots in Figures \ref{CDF}g and \ref{CDF}h. Due to their highly relativistic forward momentum ($p_{\rm xi}=80{\rm MeV/c}$), deceleration of the jet ions by the ambient plasma will take place in later stages, after the formation of LS in the ambient particles. Therefore, the thermodynamic properties of jet and ambient plasmas (density and temperature) would be different across the LS. It leads to the formation of a double layer (will be discussed in Section \ref{Formation of the double layers}) which causes tapping another fraction of the ambient ions in the right side of the CD. This population will become part of the LS.

\subsection{Evolution of the TS}\label{Evolution of the TS}

\begin{figure}[h]
\begin{center}
\includegraphics[scale=0.35]{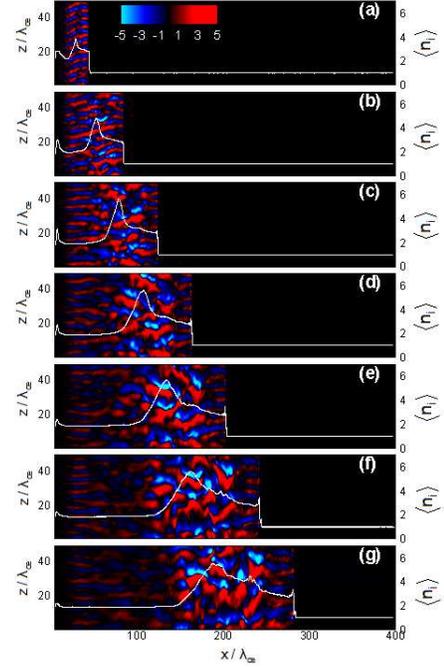}
\caption{The time evolution of the TS structure. Sequence snapshots of the magnetic field component $B_{\rm y}$ at $y=24\lambda_{\rm ce}$ from $t=40\omega_{\rm pe}^{-1}$, panel (a), up to $t=280\omega_{\rm pe}^{-1}$, panel (g), with an interval of $\Delta t=40\omega_{\rm pe}^{-1}$. Over-plotted in each panel shows the transversely averaged total ion density normalized to the density in the unshocked ambient.}
\label{RSHF}
\end{center}
\end{figure}

The continuous stream of particles jet and the inability of the particles to cross the CD result in the formation of shocks on both sides of it. Since the ambient plasma located in the right side of our simulation box represents the interstellar medium and the jet plasma coming from the left represents the ejecta, we designate the right shock as the LS and the left shock as the TS. The time evolution of the TS structure is illustrated in Figures \ref{RSHF} as a sequence of snapshots that show the magnetic field component $B_{\rm y}$ and the averaged total ion density from $t=40\omega_{\rm pe}^{-1}$ (Figure \ref{RSHF}a) up to $t=280\omega_{\rm pe}^{-1}$ (Figure \ref{RSHF}g) with an interval of $\Delta t=40\omega_{\rm pe}^{-1}$. They show that the TS propagate in the positive $x$-direction with $\beta_{\rm ts1}=0.66$. The peak value of the total ion density corresponding to the TS reaches $n_{31}/n_{41}=2.9$ at $t=280\omega_{\rm pe}^{-1}$ (Figure \ref{RSHF}g), in well agreement with the hydrodynamic jump conditions for a relativistic gas which predict $\beta_{\rm ts1}=0.68$ and $n_{31}/n_{41}=2.8$ for the TS in the ambient rest frame (see Table \ref{dsh}).

\begin{figure}[h]
\begin{center}
\includegraphics[scale=0.3]{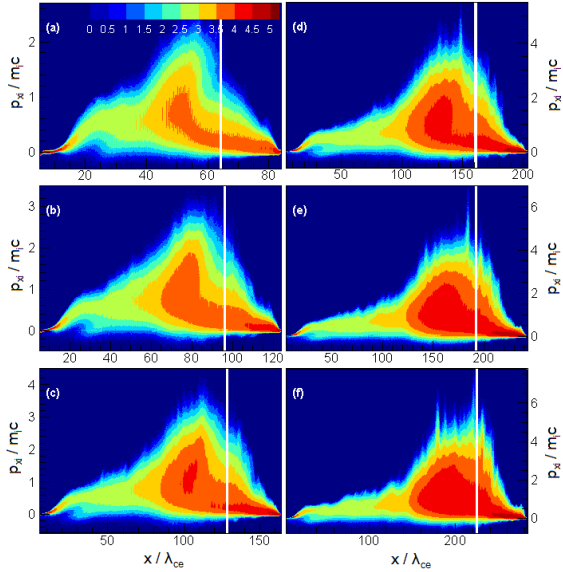}
\caption{The longitudinal phase-space distribution of ambient ions expressed in $\log[N(x,p_{\rm x})]$ during evolution of the TS structure from $t=80\omega_{\rm pe}^{-1}$, panel (a), up to $t=280\omega_{\rm pe}^{-1}$, panel (f), with an interval of $\Delta t=40\omega_{\rm pe}^{-1}$. The position of the CD at each time is shown by a vertical white line.}
\label{RSHF1}
\end{center}
\end{figure}

As shown before in Figure \ref{CDF}g, prior to the full formation of the CD, a fraction of ambient ions is present in a deeper length through the jet stream due to their higher inertia against sweeping by particles jet. They are continuously pushed towards the CD by the incoming jet stream (see Figures \ref{RSHF1}). Encountering with the CD, these ambient ions are reflected back into the left side of the CD. Therefore, the reflected ambient ions are trapped and start to pile up in the left side of the CD. This process results in the formation of the ambient ions pile in the TS structure.

\begin{figure}[h]
\begin{center}
\includegraphics[scale=0.3]{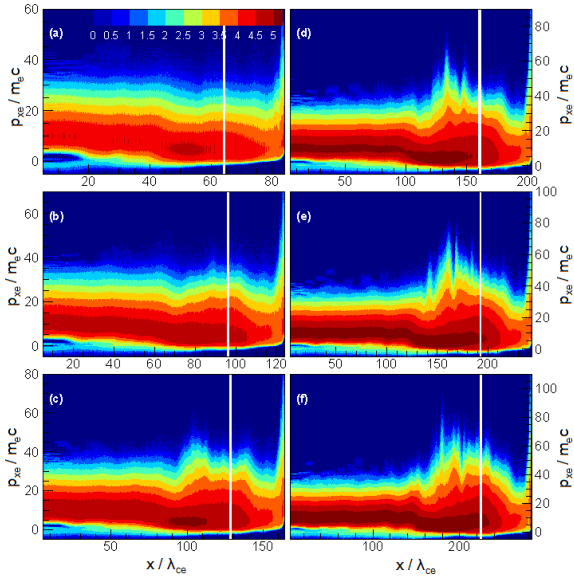}
\caption{The longitudinal phase-space distribution of jet electrons expressed in $\log[N(x,p_{\rm x})]$ during evolution of the TS structure from $t=80\omega_{\rm pe}^{-1}$, panel (a), up to $t=280\omega_{\rm pe}^{-1}$, panel (f), with an interval of $\Delta t=40\omega_{\rm pe}^{-1}$. The position of the CD at each time is shown by a vertical white line.}
\label{RSHF2}
\end{center}
\end{figure}

The electron contribution in the TS structure belongs to the jet electrons (see Figures \ref{CDF}a and \ref{CDF}b). The induced magnetic fields due to the Weibel-like instabilities in the jet-ambient collision region resist against propagating of incoming jet electrons into the ambient plasma which cause deceleration of jet electrons and formation of the CD (Figures \ref{RSHF2}). The formed CD allows no more jet electrons to pass into the ambient medium. They are effectively stopped at the left side of CD and start to pile up as a part of the TS structure.

\begin{figure}[h]
\begin{center}
\includegraphics[scale=0.3]{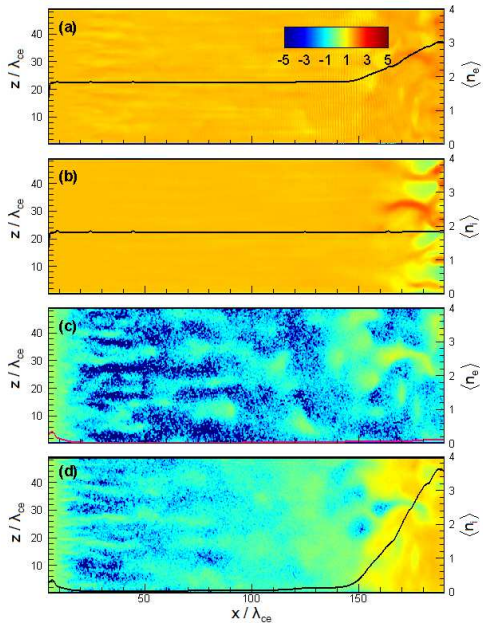}
\caption{Structure of the trailing edge at time $t=280\omega_{\rm pe}^{-1}$. The density of the particle in log scale with an over-plotted line for the average density of the particle normalized to the density in the unshocked ambient is shown for the: (a) jet electron, (b) jet ion, (c) ambient electron, and (d) ambient ion, respectively.}
\label{RST1}
\end{center}
\end{figure}

\begin{figure*}[tbph]
\begin{center}
\includegraphics[scale=0.4]{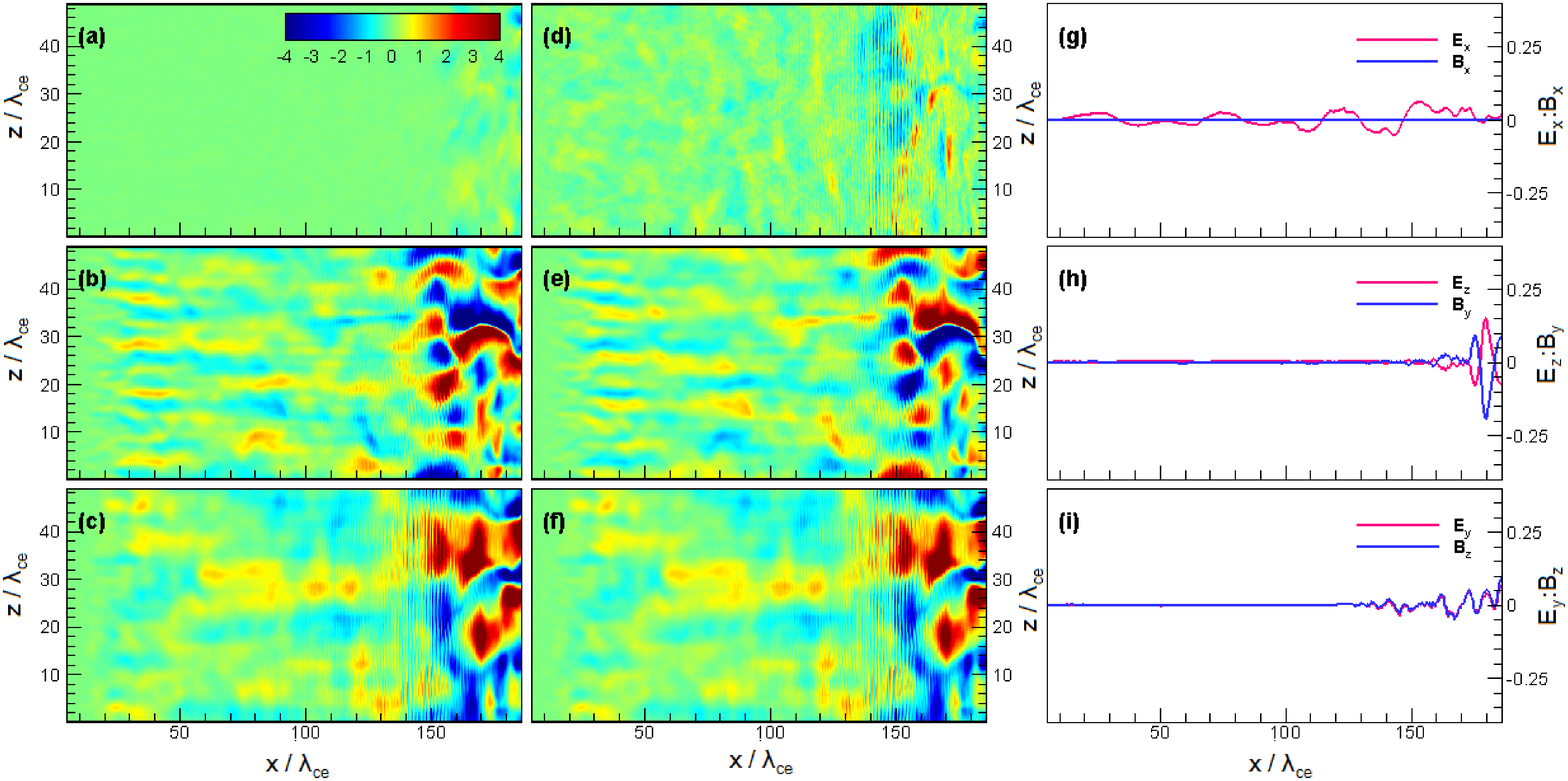}
\caption{Structure of the trailing edge at $t=280\omega_{\rm pe}^{-1}$. Panels (a), (b) and (c) show the components of the magnetic field $B_{\rm x}$, $B_{\rm y}$, and $B_{\rm z}$, respectively, at $y=24\lambda_{\rm ce}$. Panels (d), (e) and (f) show the components of the electric field $E_{\rm x}$, $E_{\rm z}$, and $E_{\rm y}$, respectively, at $y=24\lambda_{\rm ce}$. The transversally averaged field components $(E_{\rm x}: B_{\rm x})$, $(E_{\rm z}: B_{\rm y})$, and $(E_{\rm y}: B_{\rm z})$ are shown in panels (g), (h) and (i), respectively.}
\label{RST2}
\end{center}
\end{figure*}

At $t=280\omega_{\rm pe}^{-1}$, the compression ratio for the TS reaches the level of $n_{31}/n_{41}=2.9$ (see Figure \ref{RSHF}g) predicted by the hydrodynamic jump conditions for a 3D relativistic plasma with adiabatic index $\tilde{\Gamma}=4/3$. The compression of electrons is provided solely by the jet electron component (Figures \ref{RST1}a and \ref{RST1}c), while the ion contribution is supplied by the ambient ions (Figures \ref{RST1}b and \ref{RST1}d). The extended region between the unshocked and shocked jet constitutes the trailing edge. The structure of the trailing edge is exclusively controlled by the strongly nonlinear jet-ambient interactions which result in the formation and merger of current filaments due to the Weibel-like instabilities \citep{bre10}. In the vicinity of the TS, the corresponding electromagnetic fields are predominantly transverse. The transverse electric and magnetic fields are related to each other via {\boldmath${E}=-\beta\times{B}$} where {\boldmath$\beta$} is the velocity of the carriers. The carriers move roughly at the speed of light in the $x$-direction (drift velocity $v_{\rm d}=E/B \simeq c$), hence $\beta\simeq\beta_{\rm x}\simeq1$. As the results, the transverse fields are as $E_{\rm y}=B_{\rm z}$, and $E_{\rm z}=-B_{\rm y}$ (see Figures \ref{RST2}). These electric fields cause heating of the particles in transverse directions. Ahead of the filaments (toward the unshocked jet), the electrons of the ambient plasma are absent (Figure \ref{RST1}c). However, a population of hot jet electrons which has been reflected in the CD region streams with slightly relativistic velocity against the incoming jet (see Figures \ref{RSHF2}). This process excites a Weibel-like two-stream instability \citep{med99,fre04,hed04} between the reflected jet electrons and incoming jet electrons that construct a longitudinal electrostatic perturbations as $E_{\rm x}$ (Figure \ref{RST2}g) and associated density modulations, further to the filamentation of the trailing edge (see Figures \ref{RST2}). The amplitude of Weibel-like two-stream perturbations saturates at small level, and their major effect is to heat up the jet electrons in the trailing edge. Toward the TS, the longitudinal electrostatic perturbations become amplified through the stream of the reflected ambient ions in the CD region (see Figures \ref{RSHF1}), and the fluctuations in $E_{\rm x}$ are enhanced (Figure \ref{RST2}g). The longitudinal structures coexist with transverse filaments, indicating that the Weibel-like instability and the density modulation operate in parallel and propagate obliquely.


\subsection{Evolution of the LS}\label{Evolution of the LS}

\begin{figure}[h]
\begin{center}
\includegraphics[scale=0.3]{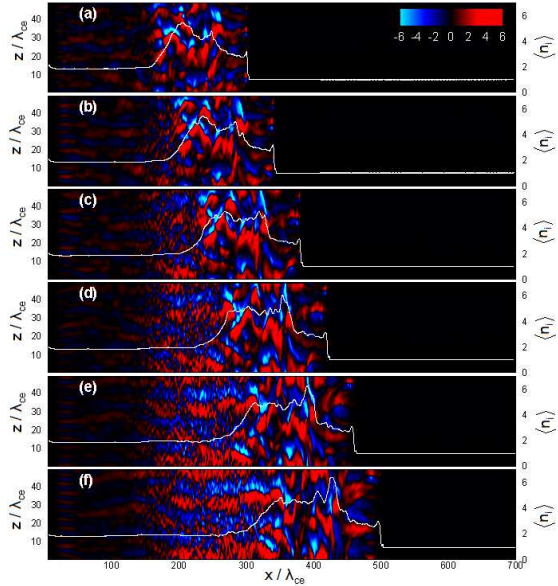}
\caption{The time evolution of the TS structure.  Sequence snapshots of the magnetic field component $B_{\rm y}$ at $y=24\lambda_{\rm ce}$ from $t=300\omega_{\rm pe}^{-1}$, panel (a), up to $t=500\omega_{\rm pe}^{-1}$, panel (f), with an interval of $\Delta t=40\omega_{\rm pe}^{-1}$. Over-plotted in each panel shows the transversely averaged (in $yz$-plane) total ion density normalized to the density in the unshocked ambient.}
\label{FSHF}
\end{center}
\end{figure}

The evolution of the LS structure is displayed in Figures \ref{FSHF} where the magnetic field component $B_{\rm y}$ and the averaged total ion density are shown in sequent snapshots from $t=300\omega_{\rm pe}^{-1}$ (Figure \ref{FSHF}a) up to $t=500\omega_{\rm pe}^{-1}$ (Figure \ref{FSHF}f) with an interval of $\Delta t=40\omega_{\rm pe}^{-1}$. As one can see, a density compression appears primarily in the ambient plasma at late stages ($t\simeq300\omega_{\rm pe}^{-1}$) that we designate as the LS. The compression ratio rises with time until reaches $n_{21}/n_1=6.5$ at the end of the simulation $t=500\omega_{\rm pe}^{-1}$ (Figure \ref{FSHF}f). The LS structure moves with a speed $\beta_{\rm ls1}=0.89$ in the positive $x$-direction. In the formed double shock structure, the CD moves in the positive $x$-direction with a speed $\beta_{\rm cd1}=0.80$. The hydrodynamic jump conditions for the LS predict $\beta_{\rm ls1}=0.85$ and $n_{21}/n_{1}=16$ in the ambient rest frame (Table \ref{dsh}). Hence, the density jump for the shocked ambient is about a factor of $\sim2.5$ smaller than theoretically predicted for a fully developed LS.

\begin{figure}[h]
\begin{center}
\includegraphics[scale=0.3]{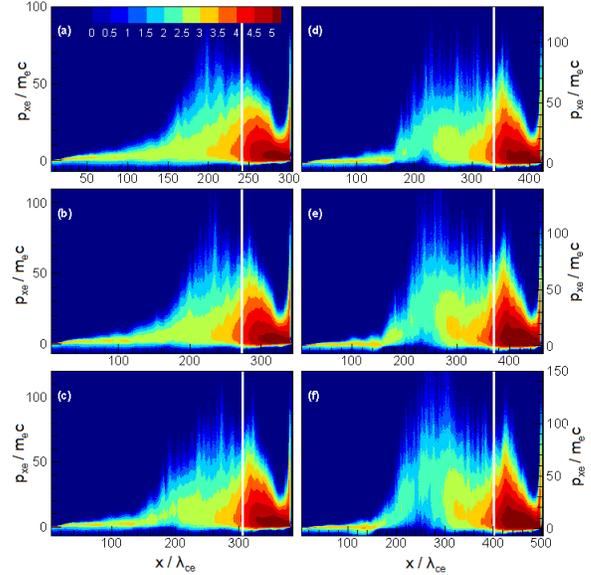}
\caption{The longitudinal phase-space distribution of ambient electrons expressed in $\log[N(x,p_{\rm x})]$ during evolution of the LS structure from $t=300\omega_{\rm pe}^{-1}$, panel (a), up to $t=500\omega_{\rm pe}^{-1}$, panel (f), with an interval of $\Delta t=40\omega_{\rm pe}^{-1}$. The position of the CD at each time is shown by a vertical white line.}
\label{FSHF1}
\end{center}
\end{figure}

\begin{figure}[h]
\begin{center}
\includegraphics[scale=0.3]{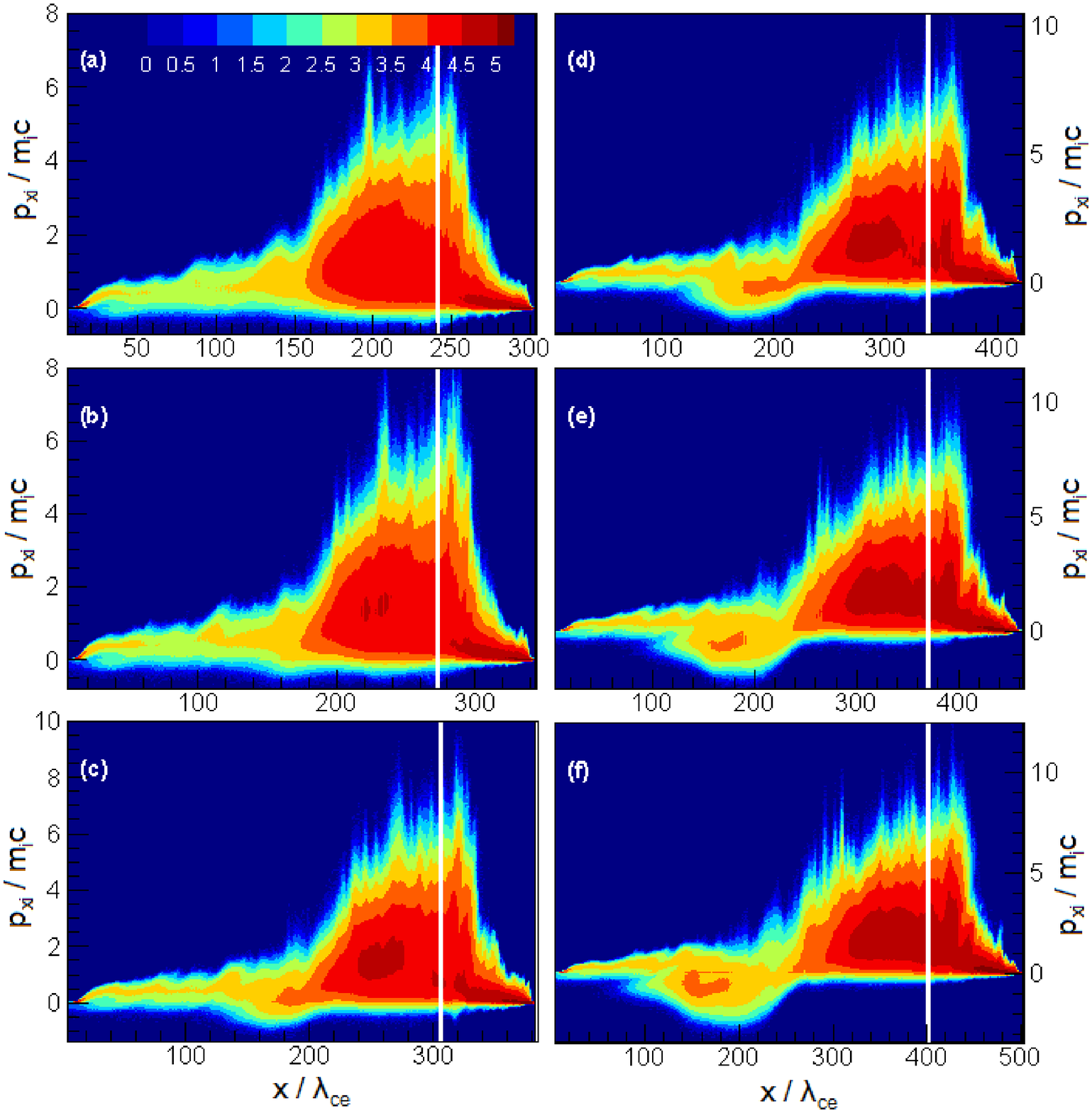}
\caption{The longitudinal phase-space distribution of ambient ions expressed in $\log[N(x,p_{\rm x})]$ during evolution of the LS structure from $t=300\omega_{\rm pe}^{-1}$, panel (a), up to $t=500\omega_{\rm pe}^{-1}$, panel (f), with an interval of $\Delta t=40\omega_{\rm pe}^{-1}$. The position of the CD at each time is shown by a vertical white line.}
\label{FSHF2}
\end{center}
\end{figure}

The ambient particles (both electrons and ions) are swept by the incoming jet stream. Due to the CD formation in early stages and reflection by the CD, the ambient electrons are mainly trapped in the right side of the CD (Figures \ref{FSHF1}) and create a compressed region as part of the LS structure. In regards to the ambient ions, as discussed in Section \ref{Evolution of the TS}, they are also present in deeper length of the trailing edge due to their higher rigidity against the incoming jet stream. On other hand, the formed CD and continuous sweeping by jet stream accumulate part of the ambient ions at the right side of the CD (Figures \ref{FSHF2}). This population contributes in the LS structure. Furthermore, during evolution of the LS, reflection of the ambient ions against the incoming jet happen which these hot counter-streaming ions can be clearly seen as a population with negative momenta in Figures \ref{FSHF2}. Counter-streaming ions play important role regarding the double layer preservation in the trailing edge which will be discussed in Section \ref{Formation of the double layers}.


\begin{figure}[h]
\begin{center}
\includegraphics[scale=0.3]{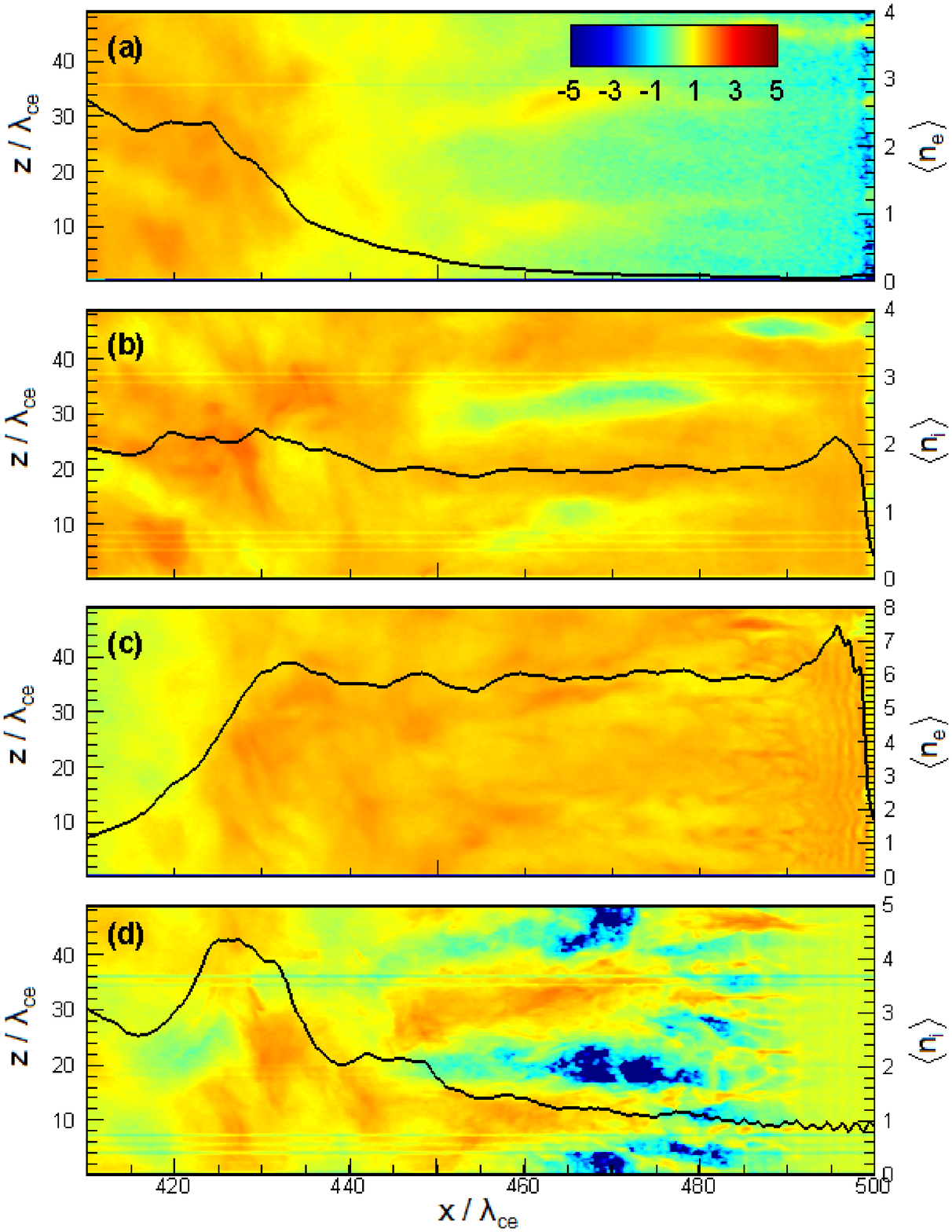}
\caption{Structure of the leading edge at time $t=500\omega_{\rm pe}^{-1}$. The density of the particle in log scale with an over-plotted line for the average density of the particle normalized to the density in the unshocked ambient is shown for the: (a) jet electron, (b) jet ion, (c) ambient electron, and (d) ambient ions, respectively.}
\label{FST1}
\end{center}
\end{figure}

By the time the simulation ends, the compression ratio for the TS has not yet reached the compression ratio of a fully developed hydrodynamic shock. The compression of electrons is supplied dominantly by the ambient electrons (Figures \ref{FST1}a and \ref{FST1}c), although the deeply penetrated jet electrons that are trapped in the right side of the CD (see Figure \ref{FST1}a beyond $x=420\lambda_{\rm ce}$) slightly contribute in the LS structure. The ion contribution is exclusively provided by the ambient ions (Figures \ref{FST1}b and \ref{FST1}d). The extended region between the unshocked and shocked ambient represents the leading edge.

\begin{figure*}[tbph]
\begin{center}
\includegraphics[scale=0.4]{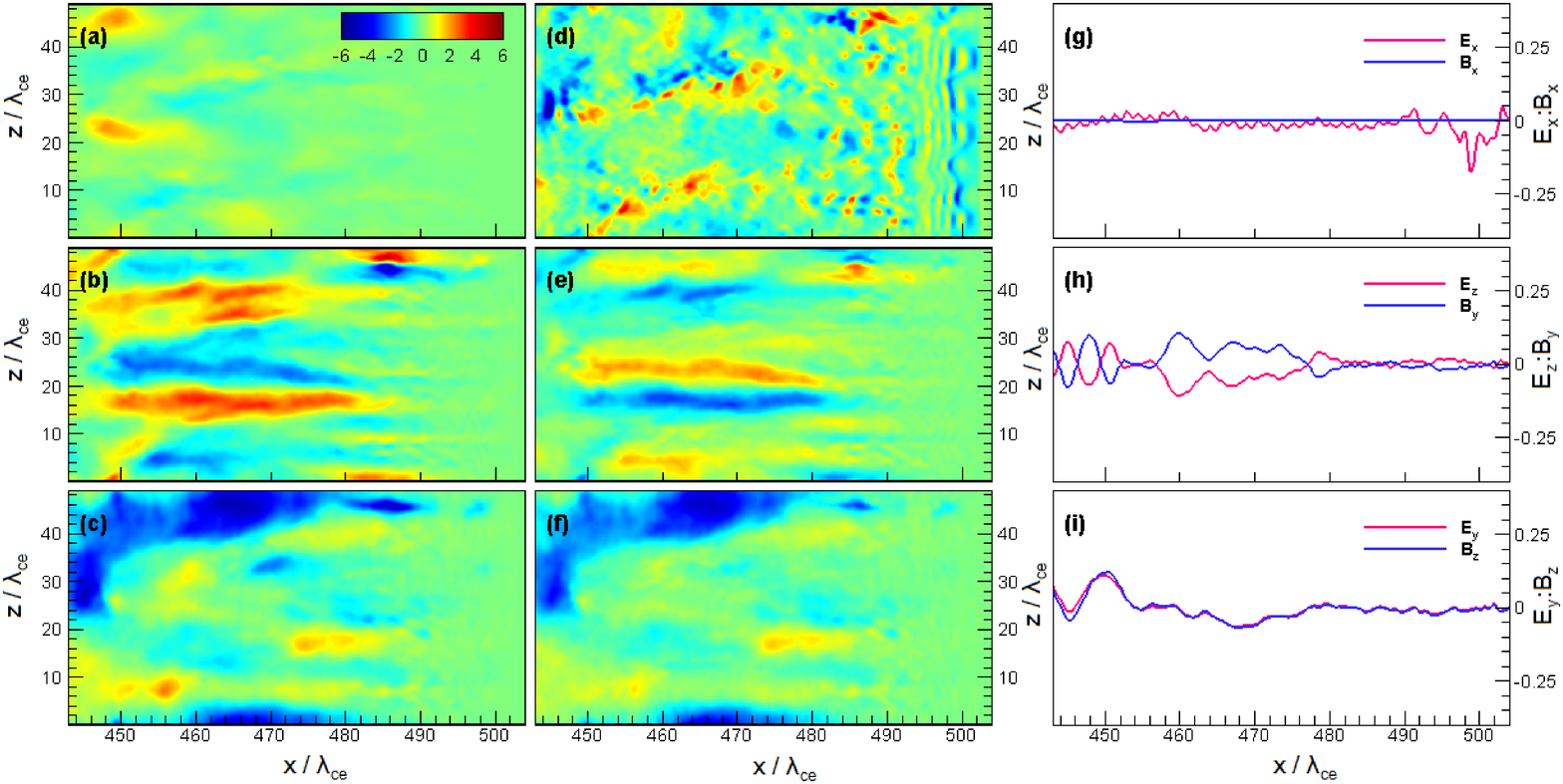}
\caption{Structure of the leading edge at $t=500\omega_{\rm pe}^{-1}$. Panels (a), (b) and (c) show the components of the magnetic field $B_{\rm x}$, $B_{\rm y}$, and $B_{\rm z}$, respectively, at $y=24\lambda_{\rm ce}$. Panels (d), (e) and (f) show the components of the electric field $E_{\rm x}$, $E_{\rm z}$, and $E_{\rm y}$, respectively, at $y=24\lambda_{\rm ce}$. The transversally averaged field components $(E_{\rm x}: B_{\rm x})$, $(E_{\rm z}: B_{\rm y})$, and $(E_{\rm y}: B_{\rm z})$ are shown in panels (g), (h) and (i), respectively.}
\label{FST2}
\end{center}
\end{figure*}

The structure of the electromagnetic fields in the leading edge is mainly controlled by the relativistic ion beam-plasma instabilities, where propagation of dense jet ions into the ambient ions excites the Weibel-like instabilities with wave vectors oriented obliquely to the jet propagation direction \citep{bre10}. The Weibel-like instabilities lead to current filamentation (see Figure \ref{FST1}d) and the generation of transverse magnetic fields (Figures \ref{FST2}a, \ref{FST2}b, and \ref{FST2}c). In contrast with the ordinary filamentation instabilities, the electric fields are not purely transverse and there is a finite electrostatic component (see Figures \ref{FST2}d, \ref{FST2}e, \ref{FST2}f and \ref{FST2}g). The relation between the transverse electric and magnetic fields is same as the trailing edge where {\boldmath${E}=-\beta\times{B}$} and hence $E_{\rm y}=B_{\rm z}$, and $E_{\rm z}=-B_{\rm y}$ (Figures \ref{FST2}). Only when the jet and ambient plasmas are quite symmetric (i.e., same density, temperature, and drift velocity), the filamentation instability would be purely transverse \citep{bre05,bre10}. In order not to result in any space charge, the beam and ambient plasmas must pinch absolutely at the same rate. However, this rate highly depends on both the thermal spread (since thermal pressure opposes the magnetic pinching) and the relativistic momentum (and thus the Lorentz factors) of the two populations. Charge imbalance hence appears whenever these parameters are different (see also \citet{cho14}). The induced magnetic fields influence the motion of particles. The jet ions are slightly decelerated in bulk and develop a population of slow particles. At the same time the ambient ions, in which filamentation is strongest (Figure \ref{FST1}d), are heated. Thick filaments in the ambient ions are surrounded by electrons. The bulk kinetic energy released by the decelerated jet ions is converted to electron heating in the electric fields that accompany the ion filaments. The volume between the filaments is depleted of ambient ions and occupied by the jet ions. The filamentary structures in the jet electrons are more diffuse on account of their higher temperature (Figure \ref{FST1}a).

\subsection{Formation of the double layers}\label{Formation of the double layers}

\begin{figure*}[tbph]
\begin{center}
\centering
\includegraphics[scale=0.4]{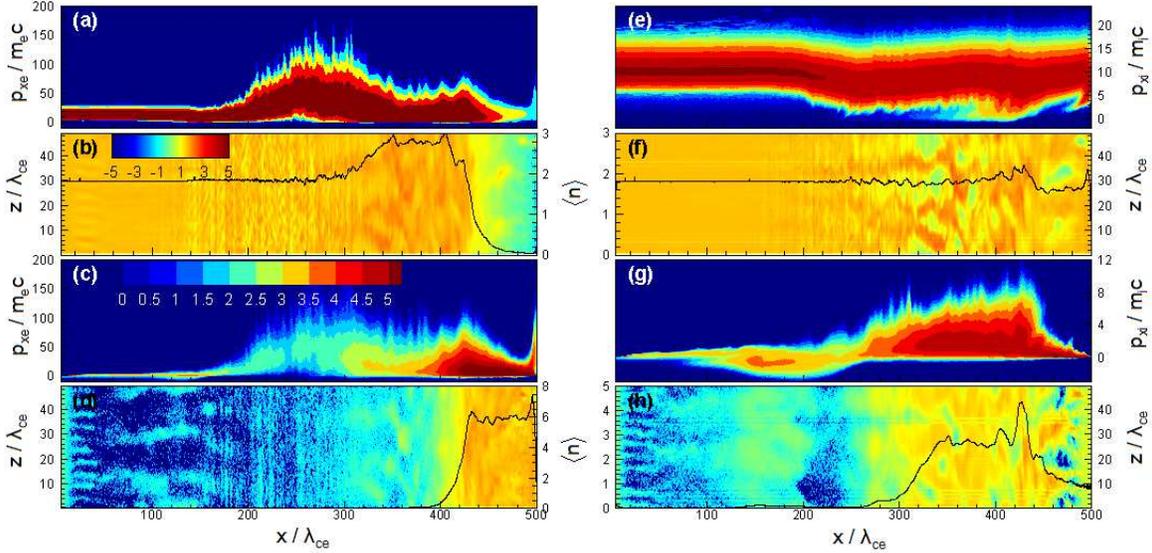}
\caption{Structure of the jet-ambient interaction at time $t=500\omega_{\rm pe}^{-1}$ when the fastest jet ions reach $x=505\lambda_{\rm ce}$. The longitudinal phase-space distribution and density in log scale are displayed for: jet electrons in panels (a) and (b), ambient electrons in panels (c) and (d), jet ions in panels (e) and (f), and ambient ions in panels (g) and (h). Over-plotted line in panel (a), (c), (e), and (g), shows the average momentum in $x$-direction. Over-plotted line in panel (b), (d), (f), and (h), shows the transversely averaged (in $yz$-plane) density normalized to the density in the unshocked ambient. In panels (a), (c), (e), and (g), the phase-space distributions are expressed in $\log[N(x,p_{\rm x})]$.}
\label{DL1}
\end{center}
\end{figure*}

As the parallel density structures advect toward the TS, the density of the ambient ion and shock-reflected ambient ions increases (see Figures \ref{FSHF2}). As a result of shock reflection, a hole in the ambient ion will be appeared within the trailing edge (see $200\lambda_{\rm ce}\lesssim x\lesssim260\lambda_{\rm ce}$ in Figure \ref{DL1}h). The hole is filled with the shocked jet electrons, small fraction of the ambient electrons those are trapped in the trailing edge due to the CD, and jet ions. This process forms a double layer plasma and the associated ambipolar electrostatic field causes trapping of the shock-reflected ambient ions behind the electrostatic field (see $x\lesssim200\lambda_{\rm ce}$ in Figure \ref{DL1}h). The electrons are locally accelerated to high energy ($p_{\rm e}\simeq75{\rm MeV/c}$ in Figure \ref{DL1}a) and convect toward the TS region. The accelerated jet electrons and the reflected ambient ions correspond to the freely streaming particle species in a double layer plasma as discussed in \citet{blo78}. Figure \ref{DL1} presents the structure of the jet-ambient interaction at time $t=500\omega_{\rm pe}^{-1}$ that is representative of the characteristics discussed above. The figure shows the longitudinal phase-space distribution of particles, and the density of particle in log scale with an over-plotted which illustrates the average density, for jet electron, ambient electron, jet ion, and ambient ions.

\begin{figure}[h]
\begin{center}
\centering
\includegraphics[scale=0.35]{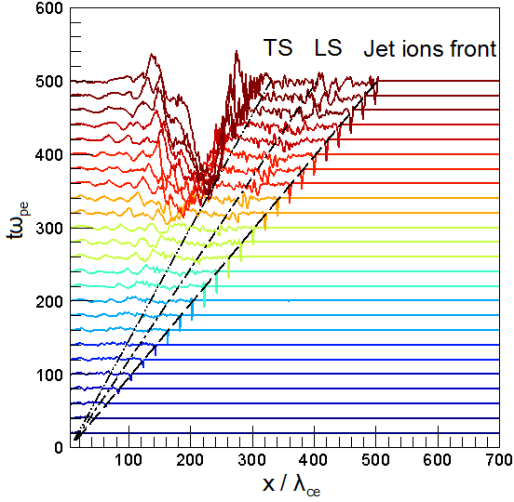}
\caption{The stacked profiles of the transversely averaged $E_{\rm x}$ is shown from  $t=20\omega_{\rm pe}^{-1}$ up to $500\omega_{\rm pe}^{-1}$ with an interval of $\Delta t=20\omega_{\rm pe}^{-1}$. Dashed-dot-dot, dashed-dot, and dashed lines represent the TS, LS, and jet ions front, respectively. }
\label{DL3}
\end{center}
\end{figure}

\begin{figure}[h]
\begin{center}
\centering
\includegraphics[scale=0.35]{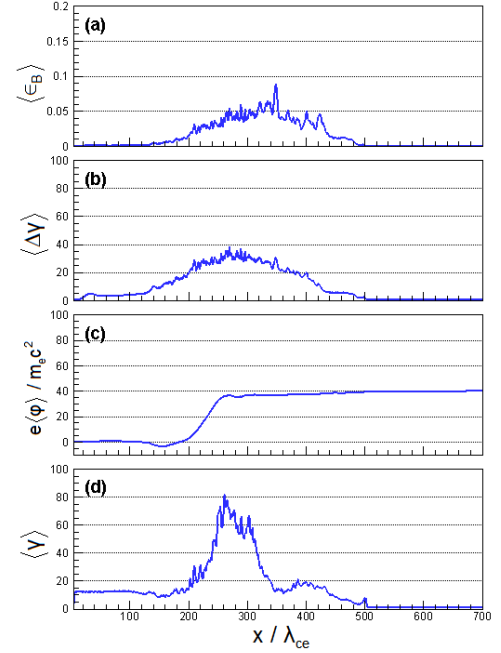}
\caption{Electrons heating and acceleration. Displayed are: (a) magnetic energy density $\epsilon_B$, normalized to the jet energy density, (b) average change in the electron energy due to the transverse electric fields of ion filaments, $\langle\Delta E_{\rm e}/m_{\rm e}c^2\rangle=\langle\Delta\gamma_{\rm e}\rangle$, (c) average change in the electron kinetic energy due to the double layer electric field, $e\langle\phi\rangle/m_{\rm e}c^2$, and (d) average electron energy, $\langle\Delta\gamma_{\rm e}\rangle$, along $x$-direction. All panels are calculated at $t=500\omega_{\rm pe}^{-1}$.}
\label{heating}
\end{center}
\end{figure}

The double layer in the trailing edge accelerates jet electrons out of the bulk to an average momentum $\langle p_{\rm e}\rangle\simeq40{\rm MeV/c}$ (see Figure \ref{DL1}a and Figure \ref{heating}d). The formed double layer is not stationary, it is one with a floating potential instead (Figure \ref{DL3}). Therefore, the energy of the jet electrons increases in time while the jet electron temperature remains unchanged (will be shown in Section \ref{Evolution of the electron distribution function}). The energy of the accelerated electrons exceeds their thermal energies, even after the Weibel-like instabilities have heated the electrons (the average $\langle p_{\rm ye}\rangle\simeq\langle p_{\rm ze}\rangle\simeq20{\rm MeV/c}$, see Figure \ref{heating}b). Therefore, the kinetic energy of the jet electrons have been increased through the double layer potential where $e\langle\phi\rangle\simeq20{\rm MeV}$ (Figure \ref{heating}c). According to the Bohm criterion \citep{blo78}, a double layer demands a drift speed that is rather faster than the thermal speed. This criterion is well satisfied whereas $v_{\rm d}\simeq c$. The accelerated jet electrons then interacts with the ambient medium through a oblique Weibel-like instability, which the corresponding electromagnetic fields are responsible for the spikes in the electrons phase-space distribution within the interval $200\lambda_{\rm ce}\lesssim x\lesssim300\lambda_{\rm ce}$ in Figures \ref{DL1}a and \ref{DL1}c. A secondary two-stream instability was also found in \citet{new01,die09}, although the jets (beams) were non-relativistic there. Principally, the type of instability is not important for the evolution of the double layer because it forms behind it. The electric field of the double layer is strong inasmuch as it can slow down the jet ions by a factor of 50 \% from the initial momentum $p_{\rm i}=80{\rm MeV/c}$ (Figure \ref{DL1}e), which supports the energy for the electron acceleration. The double layer is thus an ion decelerator which is characteristic of an electrostatic shock. The corresponding electrostatic TS involving only the ambient ions occurs at $x=340\lambda_{\rm ce}$.

Another double layer structure exists in the leading edge (see Figure \ref{DL3}) which move with a speed $\beta\simeq c$. The density and temperature of the jet and ambient plasmas differ through the leading edge (see Figures \ref{DL1}). Therefore, the quasi-neutrality is violated and a double layer will be formed. This double layer accelerates ambient electrons up to an average energy $\simeq5{\rm MeV}$ (see Figure \ref{DL1}c and Figure \ref{heating}c). Similar to the previous one, the double layer in the leading edge is strong enough to slow down the jet ions stream and supply the energy for electron acceleration.  As a results, another electrostatic shock including also the jet ions forms near the jet ions front (see Figures \ref{FST1}b and \ref{FST1}c).

The transverse electric and magnetic fields ($E\simeq B$, and the energy density stored in the magnetic fields reach about 10\% of the jet energy density in the shocked region, see Figure \ref{heating}a) in $yz$-plane result in a {\boldmath${E} \times{B}$} drift motion of electrons in $x$-direction. During this motion, the electrons become efficiently heated. The maximum attainable energy for an electron during drift motion has been estimated analytically \citep{med06,ard15} where the electron energy density is proportional to the square root of the magnetic field energy density, normalized to the total incoming energy density, $\epsilon_{\rm e}\simeq \sqrt{\epsilon_{\rm B}}$. Using this expression, the average change in the electron energy, $\langle\Delta E_{\rm e}/m_{\rm e}c^2\rangle=\langle\Delta\gamma_{\rm e}\rangle$, due to the transverse electric fields of ion filaments is displayed in Figure \ref{heating}b. As one can see, in the trailing edge, $x\lesssim340\lambda_{\rm ce}$, the electrons (mostly jet electrons) are heated by the ion filament up to $20{\rm MeV}$. Furthermore, due to the presence of a double layer in the trailing edge, the electrons can gain more energy within the double layer electric field. The maximum attainable energy through the double layer in the trailing edge is $e\langle\phi\rangle\simeq20{\rm MeV}$ (Figure \ref{heating}c). Hence, ion filaments and double layer together increase the electrons energy in the trailing edge by an average energy of $40{\rm MeV}$ (Figure \ref{heating}d). A similar process in the leading edge increases the average energy of ambient electrons energy to $5{\rm MeV}$.

\subsection{Evolution of the electron distribution function}\label{Evolution of the electron distribution function}

\begin{figure*}[tbph]
\begin{center}
\centering
\includegraphics[scale=0.4]{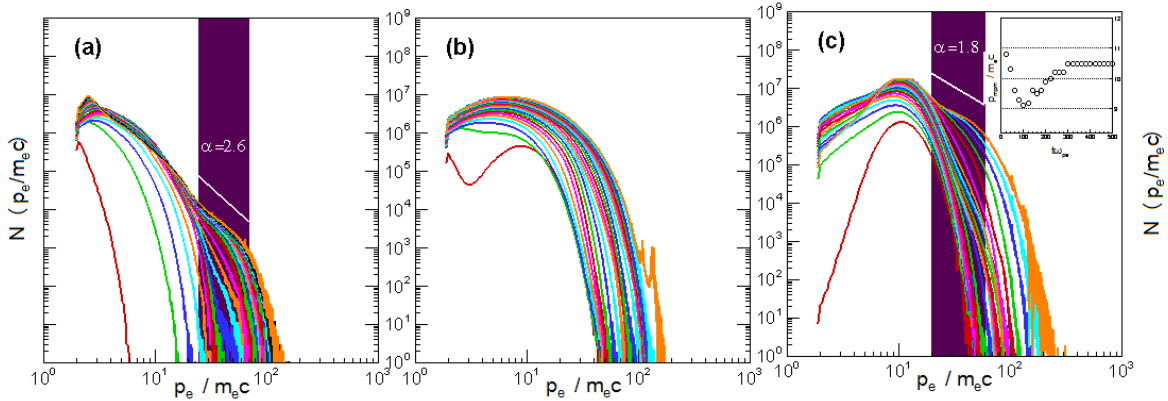}
\caption{Evolution of the electron distribution function from $t=20\omega_{\rm pe}^{-1}$ (leftmost red line) up to $500\omega_{\rm pe}^{-1}$ (rightmost orange line) with an interval of $\Delta t=20\omega_{\rm pe}^{-1}$: (a) for ambient in the leading edge taken at $(\beta_{41}-\beta_{\rm ls1})t\omega_{\rm pe}^{-1}$, (b) for ambient+jet in the shocked region taken at $(\beta_{\rm ls1}-\beta_{\rm ts1})t\omega_{\rm pe}^{-1}$, and (c) for jet in the trailing edge taken at $x/\lambda_{\rm ce}\lesssim\beta_{\rm ts1}t\omega_{\rm pe}^{-1}$. White line in panel (a) and (c) shows a power-law fit to the non-thermal component in the electron distribution function at the latest time. The inset in panel (c) shows the time evolution of the most probable momentum for jet electrons, $p_{\rm mpm}$.}
\label{spectrum}
\end{center}
\end{figure*}

The common observational characteristic of PWNe, GRBs afterglows, and AGN jets is a broad non-thermal spectrum of synchrotron and inverse Compton emission that extend from the radio up to the gamma-ray band. One of the key ingredients in creating this non-thermal spectrum is a non-thermal, high-energetic electron population. This population may be seen in the electron distribution function, where a pure 3D Maxwell-J\"{u}ttner distribution (in our case) does not account for the high energies. In fact, a more complex distribution function is expected as a result of electron acceleration. Shown in Figures \ref{spectrum} are the evolution of the electron distribution function in time (Figure \ref{spectrum}a taken in the leading edge, Figure \ref{spectrum}b taken in the shocked region, and Figure \ref{spectrum}c taken in the trailing edge). At late stages, in both leading and trailing edges, the electron distribution function consists of a drifting Maxwell-J\"{u}ttner distribution (our rest frame of reference is the ambient) and a high-energy tail. The electron distribution function in the shocked region illustrates a hot well mixed population (includes jet and ambient) with a drifting Maxwell-J\"{u}ttner distribution. The electron distribution functions in Figures \ref{spectrum}a and \ref{spectrum}c clearly develop a non-thermal tail over time. For $t\gtrsim300\omega_{\rm pe}^{-1}$, when the counter-streaming shock-reflected ions come to account and strong double layer form in the trailing edge, the electron are accelerated within the double layer. In this manner, their temperature do not changed significantly. This process is visible in the inset panel of Figure \ref{spectrum}c where the most probable momentum, $p_{\rm mpm}$, is constant for $t\gtrsim300\omega_{\rm pe}^{-1}$. The white line shows a power-law fit to the non-thermal, high-energy electron population. The power-law begins around $p_{\min}=12.5{\rm MeV/c}$ and extends to high energies with an exponential cutoff. The power-law index $\alpha$, defined in $N(p_{\rm }) \propto p_{\rm }^{-\alpha}$, has a best-fit value $\alpha=2.6$ in the leading edge, and $\alpha=1.8$ in the trailing edge. The non-thermal tail in the electron distribution function (Figure \ref{spectrum}a and \ref{spectrum}c) extends with time to higher and higher energies. It clearly demonstrates that electron acceleration is efficient and persevering in time. Regarding electron distribution function in the leading edge, at time $t=500\omega_{\rm pe}^{-1}$, the non-thermal tail for $p_{\rm }\ge12.5{\rm MeV/c}$ contains $\sim1\%$ of electrons ($\sum_{{p_{\rm e}\geq p_{\min}}}N_{\rm i}/\sum N_{\rm i}$) and $\sim8\%$ of electron energy ($\sum_{{p_{\rm e}\geq p_{\min}}}N_{\rm i}E_{\rm i}/\sum N_{\rm i}E_{\rm i}$) in the leading edge. The acceleration efficiency for electron is $\sim23\%$ by number and $\sim50\%$ by energy in the trailing edge, calculated in the same way as the leading edge.

Theoretically, an ensemble of electrons with a power-law energy distribution function $N(\gamma)d\gamma\propto{\gamma}^{-\alpha}d\gamma$ (for the ultra-relativistic speeds $\gamma\propto p$) result in a radiation spectrum $F(\nu)=\nu^{-\rm s}$ \citep{ryb79}, where the spectral index $s$ is related to the particle distribution index $\alpha$ by $s=(\alpha-1)/2$. Therefore, $\alpha=1.8-2.6$ in the electron energy distribution results in the spectral index $s=0.4-0.8$ which is in the range of the radio up to optical and X-ray emission \citep{bie97,pan01,pan02}.

\subsection{Dependence on the dimensionality}\label{Dependence on the dimensionality}

Our reference run is performed on a 3D spatial domain. To examine effect of the dimensionality, we have run a simulation with the same physical parameters as in our reference run, but in 2D computational domain.  For the 2D run, the box size along the $z$-direction is only $1.6c/\omega_{\rm pe}$ (8 grid cells). We find that the phase-space distributions of the particles and density structure agree well in terms of both the formed shock structure and the double layers in the trailing and leading edges. However the adiabatic index $\tilde{\Gamma}=3/2$ in the 2D domain results in the slower shocks ($\beta_{\rm ts1}=0.60$ and $\beta_{\rm ls1}=0.87$) and smaller particle compression ($n_{31}/n_{41}=2.0$ and $n_{21}/n_{1}=11.63$) compared to the 3D structure. 

\begin{figure}[h]
\begin{center}
\centering
\includegraphics[scale=0.32]{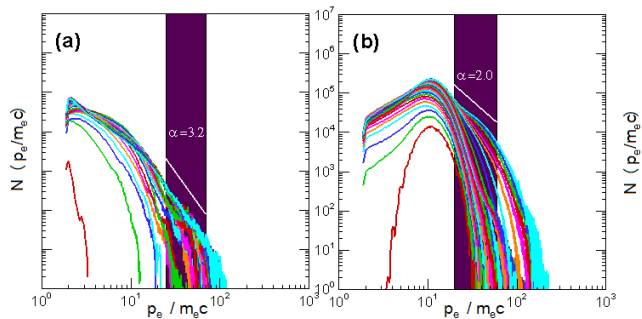}
\caption{2D run: Evolution of the electron distribution function from $t=20\omega_{\rm pe}^{-1}$ (leftmost red line) up to $560\omega_{\rm pe}^{-1}$ (rightmost cyan line) with an interval of $\Delta t=20\omega_{\rm pe}^{-1}$: (a) for ambient in the leading edge taken at $(\beta_{41}-\beta_{\rm ls1})t\omega_{\rm pe}^{-1}$, and (b) for jet in the trailing edge taken at $x/\lambda_{\rm ce}\lesssim\beta_{\rm ts1}t\omega_{\rm pe}^{-1}$. White line shows a power-law fit to the non-thermal component in the electron distribution function at the latest time.}
\label{spectrum2}
\end{center}
\end{figure}

The time evolution of the electron distribution function from the 2D run is displayed in Figures \ref{spectrum2}. In 2D run, the observed power-law index of the electron distribution function is $\alpha=3.2$ in the leading edge, and $\alpha=2$ in the trailing edge. The harder spectral index in 2D run can mean that the electron acceleration is more efficient than in 3D. Actually, the non-thermal tail in electron distribution function contains $\sim2.3\%$ of electrons and $\sim14\%$ of energy in the leading edge, and $\sim24.4\%$ of electrons and $\sim51.4\%$ of energy in the trailing edge, respectively. 

In the early phase, the fields generated in the 3D simulation are stronger than in 2D, due to the additional transverse dimension that the 3D instability can gather particles from. However, at later stages the growth of fields in the 2D simulation surpasses the 3D case \citep{sto15}. This is primarily caused by two effects: First, a 2D system has less degrees of freedom for the motion of particles; they are then more easily trapped and saturate in a larger amplitude. Second, ion current filaments can merge to larger transverse structures. This also can be followed in the 3D simulation but in longer times for larger box.

\section{Summary and conclusions}\label{Summary and conclusions}

Our motivation has been to recognize a mechanism that may accelerate electrons in the unmagnetized shock to energies, so that they can experience the DSA to ultra-relativistic energies. The observational radio synchrotron emissions from the supernova remnant shocks confirm the existence of such electrons \citep{uch07,eri11} although their origin is still unclear. The electrons are injected into the DSA, if their kinetic energies be comparable to the ion kinetic energies \citep{hos92,hos01,ama09,rey08,hil05}. It is believed that the electrons are pre-accelerated by instabilities, those are excited by ion beams in the transition region of shocks \citep{car88,hos92,hos01}. The origin of the ion beams is either the reflection of upstream ions by the shock or the leaking of downstream ions into the upstream plasma. However, Buneman instability \citep{bun58} and Weibel-like two-stream instability invoked in previous works \citep{hos01,hos02,hed04,med06,ama09} are not strong enough to inject the electrons into the DSA. They may just transfer a few percent of the ions kinetic energy to the electrons.

The present work investigates the secondary processes triggered by the Weibel-like instabilities with a 3D PIC code \citep{bun93,niem08}. The employed model of the simulation completely differs from the injection model used in several related papers \citep{hos01,hos02,spi8a,spi8b,ama09,mar09,sir11,sir13,guo14}. We have modeled an unmagnetized relativistic jet propagating into an ambient plasma. They contain ions and electrons. We have simulated the double shock system and our model is hence self-consistent. The jet moves with bulk speed $0.995c$ in the $x$-direction relative to the ambient plasma. The initial temperatures of species in the jet and ambient have been set to $46.25$ keV and $0.6$ keV, respectively, in their rest frame.

Three spatial directions have been resolved by the current PIC simulation, which implies that the wave spectrum driven by the Weibel-like instabilities propagate obliquely with respect to the jet propagation direction \citep{bre05,bre09,bre10}. Both filamentation and two-stream modes are present and operate simultaneously in electron heating. Consequently, strong fluctuations occur in the density of electrons and ambient ions that result in the formation of the double shock system. The conclusions of the work presented here can be summarized as answers to the remarked questions in Section \ref{Introduction}.

\begin{enumerate}
\item{``How does the double shock structure form in the unmagnetized jet-ambient interactions?''

At early times, a CD forms between the decelerated jet electron and the swept ambient electrons. Consequently, the jet electrons are accumulated at the right side of CD as part of TS. Additionally, a fraction of the ambient ions are located in the right side of CD cause of ion higher rigidity. They are swept by the jet continuous stream and contribute at the TS due to the reflection by the CD. Therefore, we have defined the TS as a pile of jet electrons and a fraction of ambient ions. On the other hand, the swept ambient electrons and the swept ambient ions in the right side of CD construct the LS. In a longer simulation, when the jet ions become significantly decelerated, we expect that jet ions contribute in the both TS and LS structures.} 

\item{``The shocks are characterized by magnetic or electrostatic forces?''

The electrostatic and magnetic effects are strongly activated at the same time in the captured double shock structure (a similar situation was also found for non-relativistic shocks in \citet{mat13}). The transverse magnetic fields are induced due to the Weibel-like instabilities in the jet-ambient interaction. These fields are dominantly azimuthal and associated with the ion current filaments. In the shocked region, the magnetic energy density, $\epsilon_B$, is near 10\% of jet energy density. Transverse electric fields are also present around the ion current filaments due to the density filamentation by Weibel-like instabilities. The longitudinal electrostatic fields are due to the formed double layers in the trailing and leading edges. Both electrostatic force and {\boldmath${E}\times{B}$} drift motion are important and play significant role in electron dynamics. However, in the shocked region, the induced magnetic fields facilitate energy transfer between the jet and ambient plasma.} 

\item{``What are the main mechanisms responsible for electron injection?'' 

At first, the electrons are heated up to a maximum energy density $\epsilon_{\rm e}\simeq \sqrt{\epsilon_{\rm B}}$ via {\boldmath${E}\times{B}$} drift motion. Additionally, the shock-reflected ambient ions trigger a double layer in the trailing edge which evolves consequently into an electrostatic shock. A double layer is also formed in the leading edge due to the decelerated jet ions and ambient electrons. The secondary electron energization process is associated with the electric fields of double layers. The drift speed of the free streaming particles is well in excess of the thermal one. It maintains the double layer structures in time. The substantial energy stored in the jet ions causes the electron acceleration up to 75 MeV. The double layers convert forward energy of jet ions into forward energy of electrons, without heating up the electrons. Electrons can thus be accelerated more efficiently by a double layer than by a shock because the latter spends part of flow energy for heating.} 

\item{``What is the resulting electron distribution function?''

The electron distribution function includes a non-thermal tail that contains $\sim1\%$ of electrons and $\sim8\%$ of electron energy in the leading edge, and $\sim23\%$ of electrons and $\sim50\%$ of electron energy in the trailing edge, respectively. The power-law fit to the non-thermal tail has index $\alpha=1.8$ in the trailing edge, $\alpha=2.6$ in the leading edge, respectively. These results confirm that the double layers are more efficient than shocks in electron acceleration. Based on the PIC simulations, the shocks efficiency in particle acceleration is $\sim1\%$ by number and $\sim10-15\%$ by energy \citep{spi8b,mar09,sir11,sir13,guo14}.} 

\item{``What is the effect of the dimensionality?''

In the performed 2D simulation, the power-law index in the non-thermal tail is $\alpha=3.2$ in the leading edge, and $\alpha=2$ in the trailing edge, respectively. The non-thermal tail contains $\sim2.3\%$ of electrons and $\sim14\%$ of electron energy in the leading edge, and $\sim24.4\%$ of electrons and $\sim51.4\%$ of electron energy in the trailing edge, respectively. These mean that the electron acceleration in 2D is more efficient than in 3D.}

\end{enumerate}

The present work uses a ion-electron mass ratio $m_{\rm i}/m_{\rm e}=16$. Although this low mass ratio is necessary to keep the computational costs of 3D simulations reasonable, it changes the growth rate of the unstable modes as well. In the early growth stage, when the ions are not included in the instabilities, the magnetic fields energy increases exponentially, independent of the mass ratio. However, the mass ratio effect becomes significant in the nonlinear phase. When it is small compared to the realistic one (1836), the saturation level of the magnetic field becomes higher, since the ion current filaments merge as similar as the electron ones, due to the mutual attraction between the filaments. Increasing the mass ratio will reduce the ion isotropization rate and the rate of kinetic energy exchange with electrons via the Weibel-like instabilities. Moreover, it is found that Weibel-like modes govern the high beam density regimes in the beam-plasma interactions \citep{brd10}. The domain of these modes expands as the mass ratio decreases. Consequently, the domains governed by the oblique modes shrink with decreasing the mass ratio. Therefore, our low mass ratio gives a higher importance to the Weibel-like instabilities than what they normally have.

Regarding the double layers, the electrostatic potential jumps in the trailing and leading edges are established by the electron density and temperature jumps across the shocks. These jumps are in turn decided by the shock jump conditions that do not change significantly for different ion-to-electron mass ratios. Hence, the electrostatic potentials of the double layers are independent on the mass ratio. However, increasing the mass ratio will increase the kinetic energy of the ions. The ions are thus more difficultly slowed down in the double layers, causing the slower rate of kinetic energy exchange between the ions and electrons. In this manner, the TS, LS, and CD acquire their steady-state velocity later.
\acknowledgments

We thank Masahiro Hoshino, and Takanobu Amano for valuable discussions. The work of KN is supported by NSF AST-0908010, AST-0908040, NASA-NNG05GK73G, NNX07AJ88G, NNX08AG83G, NNX08AL39G, NNX09AD16G, NNX12AH06G, NNX13AP-21G, and NNX13AP14G grants. The simulations presented here were performed on the KDK computer system at Research Institute for Sustainable Humanosphere, Kyoto University.







\allauthors

\listofchanges

\end{document}